\documentclass[twocolumn,showpacs,preprintnumbers,amsmath,amssymb]{revtex4}
\usepackage{amssymb}
\usepackage{amsfonts}
\usepackage{graphicx}
\usepackage{epstopdf}
\usepackage{amsmath}
\begin{document}
\preprint{to appear in PRB, Hase \it{et al.}}
\title[Short Title]{Ultrafast dynamics of coherent optical phonons and nonequilibrium electrons in
transition metals}
\author{Muneaki Hase$^{1}$}
\email{hase.muneaki@nims.go.jp}
\author{Kunie Ishioka$^{1,4}$}
\author{Jure Demsar$^{2}$}
\author{Kiminori Ushida$^{3}$}
\author{Masahiro Kitajima$^{1,4}$}
\affiliation{$^{1}$Materials Engineering Laboratory, National Institute for Materials 
Science, 1-2-1 Sengen, Tsukuba, 305-0047, Japan}
\affiliation{$^{2}$Department for Complex Matter, J. Stefan Institute, Jamova 39, 
Ljubljana, SI-1000, Slovenia}
\affiliation{$^{3}$RIKEN (The Institute of Physical and Chemical Research), 2-1 Hirosawa, 
Wako, 351-0198, Japan}
\affiliation{$^{4}$Graduate School of Pure and Applied Sciences, University of 
Tsukuba, 1-2-1 Sengen, Tsukuba, Ibaraki 305-0047, Japan}
\date{Received 21 Oct. 2004}

\begin{abstract}
The femtosecond optical pump-probe technique was used to study dynamics of
photoexcited electrons and coherent optical phonons in transition metals Zn
and Cd as a function of temperature and excitation level. The optical response
in time domain is well fitted by linear combination of a damped harmonic
oscillation because of excitation of coherent $E_{2g}$ phonon and a subpicosecond
transient response due to electron-phonon thermalization. The electron-phonon
thermalization time monotonically increases with temperature, consistent with the 
thermomodulation scenario, where at high temperatures the system can be well
explained by the two-temperature model, while below $\approx$ 50 K the
nonthermal electron model needs to be applied. As the lattice temperature
increases, the damping of the coherent $E_{2g}$ phonon increases, while 
the amplitudes of both fast electronic response and the coherent
$E_{2g}$ phonon decrease. The temperature dependence of the damping 
of the $E_{2g}$ phonon indicates that population decay of the coherent optical 
phonon due to anharmonic phonon-phonon coupling dominates the decay process. 
We present a model that accounts for the observed
temperature dependence of the amplitude assuming the photoinduced absorption
mechanism, where the signal amplitude is proportional to the photoinduced
change in the quasiparticle density. The result that the amplitude of the
$E_{2g}$ phonon follows the temperature dependence of the amplitude of the
fast electronic transient indicates that under the resonant condition both 
electronic and phononic responses are proportional to the change in the 
dielectric function. 

\end{abstract}
\maketitle

\section{INTRODUCTION}

Dynamics of nonequilibrium electrons and phonons in metals, semiconductors,
and superconductors have been the focus of much attention because of their
fundamental interest in solid-state physics. In particular, electron-phonon
interaction is the center of the interest for understanding the nature of
nonequilibrium electron relaxation [electron-phonon ($e-ph$) thermalization],
of dressed electrons (an electron plus its phonon cloud), of phonon
self-energy (real part corresponds to the change in phonon frequency), and of
phonon-induced electron-electron interaction (this effect gives the birth of
the Cooper pair in superconductors).\cite{Pines63}

In metals, relaxation dynamics of optically excited nonequilibrium electrons
has been extensively studied by transient-reflectivity (TR) or transient-
transmission (TT) pump-probe techniques using femtosecond
lasers.\cite{Ultrafast,Hohlfeld00,Sun93,Fann92,Fatti98} The $e-ph$
thermalization occurs on a subpicosecond time scale in metals and has been
traditionally described by the use of the two temperature model
(TTM).\cite{Anisimov74,Hohlfeld00,Hirori03,Groeneveld95} Besides the
observation of the dynamics of nonequilibrium electrons, coherent acoustic
phonon pulses have been investigated by using the pump-probe
angular deflection technique.\cite{Wright92,Wright94} The acoustic phonon 
pulse is generated through transient heating of the sample surface induced by
excitation with intense laser pulses. The response from acoustic phonon pulses
is typically observed on a several hundred picosecond time scale (or gigahertz 
frequency range) \cite{Bozovic}. Since the electron-phonon thermalization that we
are considering takes place on a subpicosecond time scale (or terahertz frequency
range),\cite{Hohlfeld00} coherent optical phonons in metals, which are
impulsively excited and decay within a few picoseconds, may play an important role
in relaxation of hot electrons.

Recently, Melnikov \textit{et al.} have observed the surface coherent optical
phonon ($\omega$ = 2.9 THz) at the surface of Gd metal using the second harmonic
generation (SHG) technique.\cite{Melnikov03} They focused their attention into
the generation mechanism of the coherent optical phonon and coupled coherent
magnon. The surface coherent optical phonon was excited by a transient charge
separation at the surface, which is similar proposal to the displacive
excitation mechanism (DECP).\cite{Zeiger92} Furthermore, Bovensiepen \textit{et
al.} \cite{Bovensiepen04} studied both surface and bulk coherent optical
phonons in Gd metal using the SHG and TR techniques, respectively. They
discussed the transient frequency shifts observed for the both phonons.
Watanabe \textit{et al.} revealed coherent surface vibration on Cs/Pt(111)
system by using the SHG technique.\cite{Watanabe04} The coherent stretching
vibration ($\omega$ = 2.2 THz) was generated by resonant impulsive stimulated
Raman scattering (ISRS),\cite{Garrett96} and the damping of the coherent
surface vibration was dominated by pure dephasing caused by hot electrons at
absorbate. Thus, the SHG technique has mainly enabled studying surface coherent
optical phonons on metallic surfaces. Nevertheless, investigations of bulk
coherent optical phonons in metals using conventional TR or TT pump-probe
techniques are still few mainly because of the very short optical penetration depth in
metals stemming from the absorption by free electrons (Drude absorption).
Therefore the change in the refractive index associated with the coherent
phonon oscillations are very weak in the bulk, making it very difficult to
measure. Although study of the surface coherent optical phonon in metal surfaces
offers interesting applications, e.g., controlling chemical reactions on metal
surfaces, there is also tremendous motivation to observe and control the bulk
coherent optical phonon in metals in order to study transient melting as a
precursor of phase transition\cite{Guo00}, as well as controlling
electron-phonon scattering.\cite{Wehner98}

In absorbing media, e.g., in semimetals, resonant ISRS has been proposed 
as the main generation mechanism of the coherent A$_{1g}$
phonon.\cite{Garrett96,Stevens02}  Stevens
\textit{et al.} considered both the standard Raman susceptibility and the
electrostrictive tensor as the driving force for resonant
ISRS.\cite{Stevens02} Since electrostrictive force is proportional to the
dielectric function, the amplitude of the coherent phonon depends on the
incident photon energy.\cite{Stevens02} In some metals, interband transition
occurs in the near infrared region, which appears as a peak in the imaginary part
of the dielectric function, and the interband transition could enhance the
amplitude of the coherent phonons through the resonant ISRS process. Therefore the
realization of generation and observation of bulk coherent optical phonons in
metals using the TR pump-probe technique requires (i) existence of $k$ = 0 optical
phonon, which can be observed by Raman scattering and (ii) a spectral peak in
the imaginary part of the dielectric function, which is related to interband
transition. Appropriate candidates of the metallic samples that meet the
above requirements, are Zn, Cd, and Mg, in which $k$ = 0 optical phonon modes
were, in fact, observed in Raman scattering.\cite{Grant73}

In this paper, we report on the observation of both bulk coherent optical
phonon and nonequilibrium electrons in Zn and Cd using a femtosecond
pump-probe reflectivity technique with high sensitivity of $\Delta R/R$ =
10$^{-7}$. The coherent $E_{2g}$ phonon, whose oscillatory frequency is in 
the terahertz region, is generated by optical excitation of nonequilibrium electrons
(interband transition). The amplitude of the oscillations decreases
dramatically as the temperature is raised from 7 K to room temperature.
Importantly, the temperature dependence of the amplitude of oscillation
closely follows the temperature dependence of the subpicosecond electronic transient,
suggesting that under the resonant condition both electronic and phononic 
responses are proportional to the change in the dielectric function. The relaxation 
time of the fast electronic response decreases with decreasing temperature, 
showing saturation at temperatures below 100 K. In fact, this behavior closely follows
the temperature dependence of the electron-phonon thermalization time observed
in simple metals, such as Au, Ag \cite{Groeneveld95}. In order to explain the
peculiar temperature dependence of the amplitude of the electronic transient,
we have derived a simple model, where the photoinduced change in reflectivity
is governed by photoinduced absorption, where the initial states lie near the
Fermi energy. Assuming the so-called two-temperature model \cite{Groeneveld95}
and energy conservation law we were able to reproduce the temperature
dependence of the transient amplitude extremely well - further supporting the
association of a fast transient with electron-phonon thermalization and 
correlation between electronic and phononic responses via the dielectric function. 
We have extended the range of
pump fluences up to 6 mJ/cm$^{2}$ and study phonon softening as a spectator
of thermal expansion. In Sec. V, we discuss the temperature dependence of 
the amplitude of the coherent optical phonons in Zn in terms of 
Stevens' model.\cite{Stevens02}

\section{EXPERIMENTAL TECHNIQUE}

\begin{figure}[ptb]
\includegraphics[width=8.8cm]{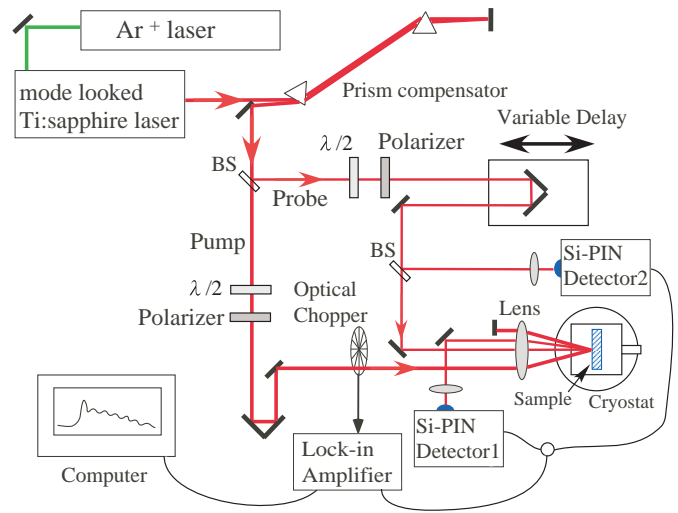}
\caption{(color online) Schematics of femtosecond pump-probe experiment. 
BS: beam splitter, $\lambda$/2: half wave-plate. The difference of the photocurrent 
of the two Si-PIN detectors was amplified, and thus reduced current noise from the 
laser source.
}
\label{Fig1}
\end{figure}
The samples used here were single crystals of Zn and Cd with cut and polished
(0001) surfaces. The femtosecond pump-probe measurements (Fig. 1) were carried
out in a temperature range from 7 to 295 K using a cryostat. The light
source used was a mode-locked Ti:sapphire laser whose pulse width was $\sim$
20 fs as determined by autocorrelation of $\sim$ 30 fs,\cite{Note1} and the
repetition rate of 87 MHz. The pump-beam power was varied from 20 to 120 mW, while the
probe-beam power was fixed at 5 mW. The polarizations of the pump and probe
beams are orthogonal to each other to avoid picking the scattered pump beam on
the detector. We used a lock-in detection where the pump-beam was mechanically
chopped at 2 kHz. Both pump- and probe-beams were focused onto a diameter of
$\sim$ 100 $\mu$m on the sample. The penetration depth of the laser light with
a wavelength of 800 nm (= 1.55 eV) was estimated to be $\sim$ 13 nm based on
the absorption coefficient of 7.59$\times$10$^{5}$ cm$^{-1}$. Therefore the
contribution to the signal from the surface oxide layer is negligibly small.
At the pump fluence of 9.2$\mu$J/cm$^{2}$ (60 mW) we estimated the maximum
electron temperature rise $\Delta T_{e}^{m}$ (assuming a thermalized electron
gas) to be $\leq$ 100 K at 7 K.\cite{Note2} Thus the experiments can be
considered as being in the reasonably weak excitation regime.\cite{Hirori03,Arbouet03} 
the TR signal ($\Delta R/R$) was measured to extract the relaxation dynamics of
nonequilibrium carriers and the bulk coherent optical phonon as a function of the
time delay after excitation pulse.\cite{Dekorsy00}

\section{GENERATION OF COHERENT OPTICAL PHONONS}

Traditionally, generation of coherent optical phonons has been described by
either DECP\cite{Zeiger92} (for absorbing media) or ISRS\cite{Yan85} (for
transparent media). In DECP mechanism, photoexcitation induces 
changes in the electronic energy distribution function, and as a result, the crystal 
lattice begins to oscillate around the new equilibrium position; $Q_{0}(t)=\kappa n(t)$, 
where $\kappa$ is a constant and $n(t)$ is the photoexcited carrier
density.\cite{Zeiger92} In the first order only the A$_{1g}$ totally symmetric
modes are coherently excited by the DECP mechanism. Recently, Stevens \textit{et
al.} have suggested that the combination of DECP and ISRS is possible in
semimetals. We briefly introduce their model below.\cite{Stevens02} 

The impulsive stimulated Raman scattering equation of motion for an optical
vibrational mode whose normal coordinate is labeled $Q$ is given
by\cite{Stevens02,Yan85}
\begin{equation}
\frac{\partial^{2}Q}{\partial t^{2}}+2\Gamma\frac{\partial Q}{\partial
t}+\Omega_{0}^{2}Q=F(t)\equiv\frac{Nv_{c}}{2}\sum_{k,l}R_{kl}E_{k}E_{l},
\end{equation}
where $\Gamma$ and $\Omega_{0}$ are the vibrational damping and 
the frequency, respectively, $N$ is the number of cells, $v_{c}$ is the volume of the unit
cell, while $E_{k,l}$ are components of the electric field delivered by the
broad spectrum of the ultrashort pump light. According to wave-vector and
energy-conservation rules, a phonon mode with frequency $\omega_{l}-\omega
_{k}=\Omega_{0}$ and wave-vector $k_{l}-k_{k}=k_{0}$ can be
excited.\cite{Dekorsy00,Yan85} The Raman tensor $R_{kl}$ denotes both the
standard Raman tensor $\chi_{kl}^{R}$ and the electrostrictive force $\pi
_{kl}^{R}$. Equation (1) is valid only if $R_{kl}$ does not depend on frequency. In
absorbing media, such as semimetals and metals, the Raman tensor $R_{kl}$ depends
on frequency and $F(t)$ becomes a function of the frequency. 
According to the theoretical work for the electrostrictive force $\pi_{kl}^{R}$ as the
generation mechanism, assuming that the dielectric function $\varepsilon
(\omega)$ varies slowly within the spectral width of the pump pulse, the
Fourier component of the driving force $F(t)$ is expressed as
follows:\cite{Stevens02},
\begin{equation}
F(\Omega)\propto\Bigl[\frac{dRe(\varepsilon)}{d\omega}+2i\frac{Im(\varepsilon
)}{\Omega}\Bigr]\int_{-\infty}^{+\infty}e^{i\Omega t}|E(t)|^{2}dt.
\end{equation}
Here $E(t)$ is the electric field of the pump pulse, i.e., $|E(t)|^{2}$
corresponds to the intensity of the pump light $|E(t)|^{2}=I(t)$. The real
part of $\varepsilon$ leads to $F(t)\propto|E(t)|^{2}$, which is impulsive and
gives Q(t)$\propto$ sin$\Omega_{0}$t, whereas the imaginary part of
$\varepsilon$ leads to $F(t)\propto\int_{-\infty}^{t}|E(t^{\prime}
)|^{2}dt^{\prime}$, which is displacive in character and gives Q(t)$\propto$
cos$\Omega_{0}$t. Since $|d$Re($\varepsilon$)/$d\omega|$$\ll$Im($\varepsilon
$)/$\Omega_{0}$ with $\Omega_{0}$ being the phonon frequency, imaginary part
of the dielectric function dominates the driving force. Thus we expect that
the coherent optical phonon generated through resonant ISRS exhibits a
displacive behavior. Solving Eq. (1) for the undamped harmonic condition
($\Gamma$ = 0) yields the coherent phonon amplitude,\cite{Stevens02}
\begin{equation}
Q_{0} \approx \frac{Im(\varepsilon)Nv_{c}\Xi_{0}}{4\pi\hbar\Omega
_{0}^{2}}\int_{-\infty}^{+\infty}e^{i\Omega_{0} t}|E(t)|^{2}dt,
\end{equation}
where $\Xi_{0}$ is the deformation potential. According to Eq. (3), the
amplitude of the coherent phonon depends on the imaginary part of the
dielectric function as well as the pump intensity $I(t)$ if the temperature is
constant, so that thermal expansion and anharmonicity of the crystal lattice,
which result in modification of $v_{c}$ and $\Omega_{0}$, are negligible.

\section{EXPERIMENTAL RESULTS AND ANALYSIS}

\begin{figure}[ptb]
\includegraphics[width=8.4cm]{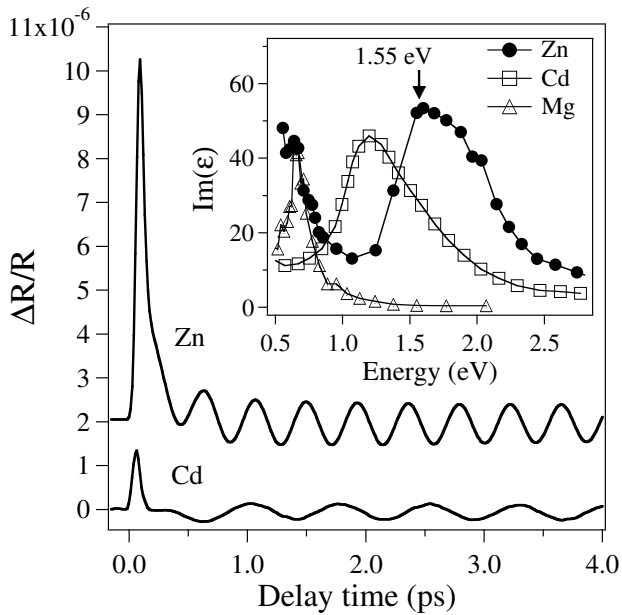}
\caption{Transient reflectivity change obtained for Zn and Cd at the lattice 
temperature 7 K. The inset shows the imaginary part of the dielectric function 
Im($\varepsilon$) for Zn (closed circles), Cd (open squares), and Mg 
(open triangles) (Ref. 27). }
\label{Fig2}
\end{figure}
Figure 2 shows the time-resolved TR signal ($\Delta R/R$) obtained for
Zn and Cd at 7 K with the pump fluence of $F_{pump}$ = 9.2$\mu$J/cm$^{2}$. The
response of the reflectivity change consist of two components. One is the
initial transient nonoscillatory response because of excitation and relaxation of
nonequilibrium electrons (fast electronic response), 
which decays in a few hundred femtoseconds. Since
the interband electronic transition near the L-point occurs at around 800 nm (1.55
eV),\cite{LB} this interband transition dominates generation of nonequilibrium
electron distribution in Zn.

The second component is the oscillatory signal due to generation of the
coherent lattice vibration (coherent phonon response). 
The time period of the observed coherent
oscillation is several hundred femtoseconds, corresponding to the bulk
$E_{2g}$ optical phonon mode as discussed below. It is to be noted that in 
addition to Zn and Cd samples, we tried to measure coherent optical phonons 
also in alkaline-earth metal Mg (0001), however we observed only very weak 
electronic response and no oscillations in the TR signal (not shown). 
The amplitude of the coherent 
optical phonon is significantly larger in Zn than that in Cd. This suggests
that the interband electronic transition, which contributes to the imaginary
part of the dielectric function Im($\varepsilon$) as shown in the inset of
Fig. 2, governs the excitation of the coherent optical phonon, although
according to Eq. (3) the phonon amplitude depends also on the deformation
potential. In fact, at the laser energy of 1.55 eV, Im($\varepsilon$) shows a
dominant peak for Zn, whereas it shows a weak shoulder for Cd. For Mg, however,
Im($\varepsilon$) is almost zero, supporting the suggestion that this
mechanism is indeed the main driving force.

Hereafter we will focus on dynamics of the coherent phonon response and
fast electronic response in Zn because the signal amplitude is much larger 
than that in Cd and therefore it is possible to make precise analysis of the time
domain signal.

\subsection{Coherent phonon response}

\begin{figure}[ptb]
\includegraphics[width=8.4cm]{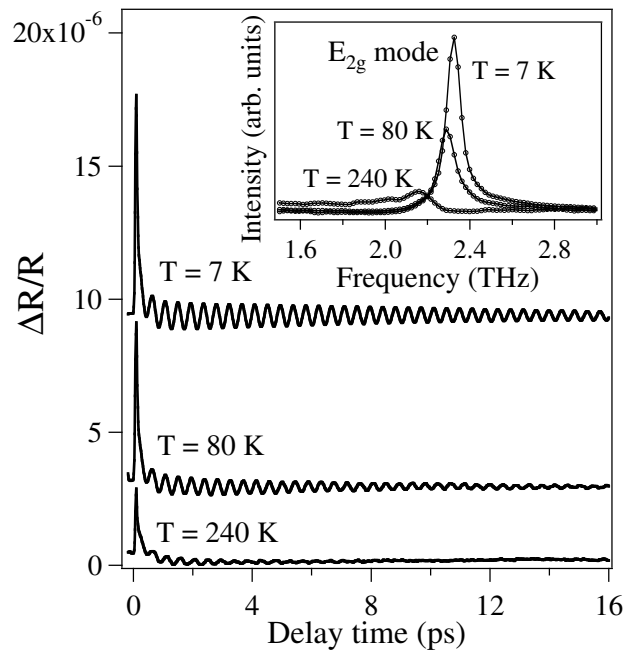}
\caption{Transient reflectivity change obtained in Zn at various lattice temperatures. 
The inset shows the Fourier transformed spectra of the time domain data. }
\end{figure}
Figure 3 shows the time-resolved TR signal ($\Delta R/R$) obtained on Zn at
various temperatures at the constant pump fluence of 9.2$\mu$J/cm$^{2}$. 
The frequency
of the coherent optical phonon observed in the Fourier transformed (FT) spectra
(see inset of Fig. 3) is 2.32 THz at 7 K, being in excellent agreement with
that of the bulk $E_{2g}$ mode observed by Raman scattering.\cite{Grant73,Schulz76} 
The Fourier transformed spectra exhibit a redshift of the peak frequency and
broadening of the line width of the $E_{2g}$ mode as the temperature increases. 
The coherent phonon signal in time domain, not including ultrafast electronic
response appearing at $t\leq$ 600 fs, was fitted by a damped harmonic oscillator; 
$Ae^{-\Gamma t}\cos(\omega_{E_{2g}}t+\phi_{0})$, 
and in this way the amplitude $A$, the decay rate (damping) $\Gamma$, the frequency 
$\omega_{E_{2g}}$, and the initial phase $\phi_{0}$ are extracted. 

\subsubsection{Temperature dependence of damping and frequency of the 
$E_{2g}$ mode}

In general, damping of the coherent phonon is governed by population 
decay (inelastic scattering) and/or pure dephasing (elastic scattering). 
In semimetal and semiconductor crystals, the decay 
process of the coherent phonon is dominated by the population decay due to 
anharmonic phonon-phonon coupling, rather than pure 
dephasing.\cite{Vallee94,Hase98} The anharmonic decay rate  
depends strongly on the lattice temperature, while the pure dephasing, e.g., 
$e-ph$ scattering, depends in metals on the hot electron density.\cite{Watanabe04} 
In order to examine the decay process of the coherent optical phonon, temperature 
dependence of the decay rate at the constant pump fluence, i.e., constant 
hot electron density, is measured. 

\begin{figure}[ptb]
\includegraphics[width=8.0cm]{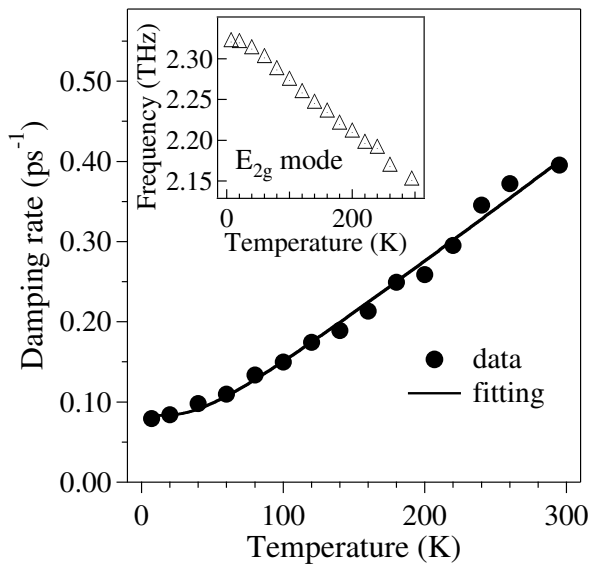}
\caption{Decay rate of the $E_{2g}$ mode together with the frequency (inset) as a 
function of the lattice temperature. The solid line represents the fit to the data 
using Eq. (4). }
\label{Fig4}
\end{figure}
Figure 4 shows the decay rate of the coherent optical phonon as a
function of the lattice temperature. The decay rate increases upon increasing
the temperature. This behavior is well explained by the anharmonic decay
model,\cite{Vallee94} in which the optical phonon decays into the two acoustic
phonons with half the frequency of the optical mode and with opposite
wave-vectors,\cite{Vallee94,Hase98}
\begin{equation}
\Gamma=\Gamma_{0}\Bigl[1+\frac{2}{exp(\frac{\hbar\Omega_{0}/2}{k_{B}T}
)-1}\Bigr].
\end{equation}
Here $\Gamma_{0}$ is the effective anharmonicity as the fitting parameter and
$k_{B}$ the Boltzmann constant. $\Gamma_{0}$ is determined to be 0.06
ps$^{-1}$. The good agreement of the time domain data with the anharmonic
decay model indicates that the damping of the coherent $E_{2g}$ mode in Zn
is due to anharmonic phonon-phonon coupling (population decay). 
The frequency of the $E_{2g}$ mode decreases as the temperature
increases as shown in the inset of Fig. 4. This temperature dependence is
qualitatively in good agreement with the anharmonic frequency shift observed
by Raman scattering measurements.\cite{Schulz76} Such a frequency shift due to
the lattice anharmonicity was also observed in III-V semiconductors, which was
reproduced by \textit{ab initio} calculations including various anharmonic
contributions (thermal expansion, third-, and fourth-order
anharmonicity).\cite{Debernardi00}

\subsubsection{Temperature dependence of the amplitude of the $E_{2g}$ mode}

\begin{figure}[ptb]
\includegraphics[width=8.8cm]{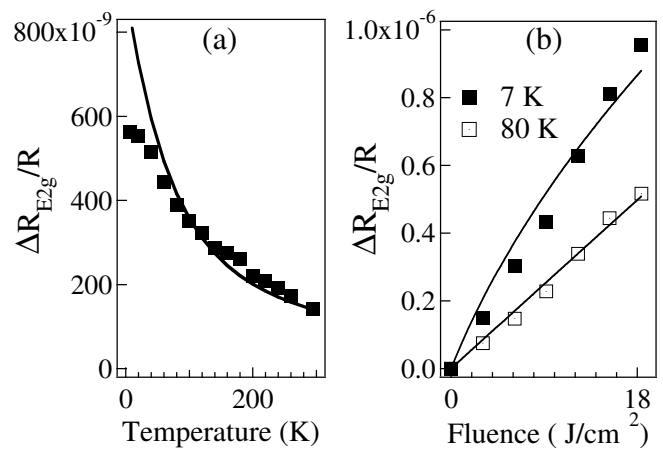}
\caption{(a) Amplitude of the coherent $E_{2g}$ phonon as a function of 
lattice temperature. (b) Amplitude of 
the coherent $E_{2g}$ phonon vs the pump fluences at the two typical
temperatures. The solid curves are the fit to the data using Eq. (12). }
\label{Fig5}
\end{figure}
As shown in Fig. 5 (a), the amplitude of the coherent $E_{2g}$
phonon significantly decreases upon increasing the lattice temperature.
The decrease in the
amplitude of the coherent optical phonon at higher temperatures is contrary to
the temperature dependence of the Raman intensity,\cite{Grant73,Weber00} confirming 
that the amplitude of the nonequilibrium coherent phonon is not determined by 
Bose-Einstein statistics $n_{k}=1/[exp(\hbar\Omega_{0}/k_{B}T)-1]$, which shows an 
increase in the phonon occupation number at higher temperatures. 
The very close behavior of the temperature dependence of the phonon and
electronic [see below in Fig. 9 (a)] responses suggests 
that the amplitudes of the fast electronic and the coherent phonon responses 
have the same origin. We will discuss the temperature dependence of the 
amplitude of the $E_{2g}$ mode in Sec. V.

\subsubsection{The photoexcitation dependence}

The photoexcitation intensity dependence of the coherent $E_{2g}$
mode was investigated in the range up to $I$ = 18.4$\mu$J/cm$^{2}$. The
amplitude monotonically increases with the pump fluence both at 7 and 80 K
as shown in Fig. 5 (b). On the other hand, the frequency and the damping of the
coherent $E_{2g}$ mode were independent on the pump fluence in this fluence
range. The latter result shows that the damping of coherent $E_{2g}$ phonons
in Zn is dominated by anharmonic phonon-phonon coupling, which is determined
by the lattice temperature, rather than by pure dephasing via scattering by 
hot electrons.\cite{Watanabe04} 

\begin{figure}[ptb]
\includegraphics[width=8.0cm]{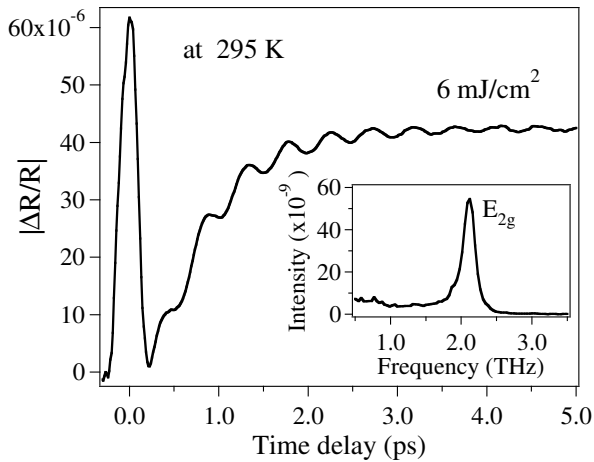}
\caption{Transient reflectivity change observed at 18 mJ/cm$^{2}$ at 295 K. 
The inset shows FT spectrum of the time domain data, showing the coherent 
$E_{2g}$ mode. }
\label{Fig6}
\end{figure}
When increasing the pump fluence by three orders of magnitude (up to 6
mJ/cm$^{2}$) using the amplified femtosecond laser system ( pulse duration of
130 fs) the time domain signal drastically changes. Figure 6 shows transient
reflectivity change observed at 6 mJ/cm$^{2}$. The width of the initial
electronic transient response is different from those observed in Figs. 2 and 3 
because the pulse duration used in Fig. 6 is much longer than that in Figs. 2 and 3.
Importantly, although the pump fluence increases by $\sim$ 10$^{3}$, the
electronic response increases only by one order of magnitude, suggesting 
some kind of a saturation behavior.\cite{DeCamp01} The frequency of the 
coherent $E_{2g}$ mode in the FT spectra (the inset of Fig. 6) is 2.10
THz, which is slightly redshifted from that obtained at 9.2$\mu$J/cm$^{2}$ at
295 K, 2.15 THz. This redshift would originate from electronic softening of
the lattice\cite{Hunsche95} or thermal expansion due to lattice
anharmonicity\cite{Perner00,Hase02} under high-density photoexcitation. We note that
strong background signal from lattice heating arises at $t>2$ ps and therefore the effect 
of thermal expansion may dominate the redshift of the coherent $E_{2g}$ mode
in Zn.

\subsection{Fast electronic response}

\begin{figure}[ptb]
\includegraphics[width=8.4cm]{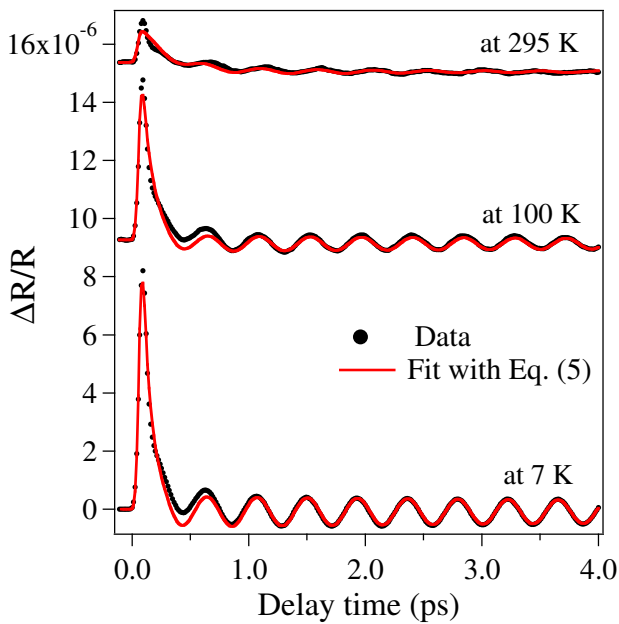}
\caption{(Color online) Transient reflectivity change
observed for Zn at 7, 100, and 295 K. The closed circles are experimental
data, and the solid curve is the fitting with Eq. (5). }
\end{figure}
The subpicosecond electronic response measured at several temperatures is
presented in Fig. 7. In order to fit the time domain data and to obtain the
amplitude and relaxation time of the fast electronic transient, we utilize a linear
combination of a damped harmonic oscillator and a single exponential decay function

\begin{equation}
\frac{\Delta R(t)}{R} = H(t)[Ae^{-\Gamma t}\cos(\omega_{E_{2g}}t+\phi_{0})+Be^{-t/\tau
_{q}}+const.]. \label{Fit}
\end{equation}
Here $H(t)$ is the Heaviside function convoluted with Gaussian to account for
the finite time-resolution, while $B$ and $\tau_{q}$ are the amplitude and relaxation 
time of the fast electronic transient, respectively. In simple
metals, such as Au, Ag,\cite{Groeneveld95} ultrafast transient is usually
attributed to the electron-phonon thermalization, which is treated by the so-
called two-temperature model (TTM). The main idea of the TTM is that because of 
the fact that electron-electron thermalization is much faster than
electron-phonon thermalization, electrons quickly (in tens of femtoseconds) thermalize 
to a temperature $T_{e}$, which can be much higher than the lattice temperature
$T_{l}$, resulting in Fermi-level smearing. In the next stage, electrons
thermalize with lattice in a characteristic electron-phonon thermalization
time, which is in metals typically in the 100 fs - 1 ps range. Since after
tens of femtoseconds, when electrons have already thermalized, the changes in
the occupied electronic density of states are limited to energies of
$k_{B}T_{e}$ near Fermi level, the photoinduced reflectivity dynamics tracks
the time-evolution of the electronic temperature. The TTM is given by the set
of two coupled heat equations \cite{Kaganov,Allen}
\begin{align}
C_{e}(T_{e})\frac{\partial T_{e}}{\partial t}  &  =-g(T_{e}-T_{l})+S(z,t),\\
C_{l}\frac{\partial T_{l}}{\partial t}  &  =g(T_{e}-T_{l}),
\end{align}
where $C_{e}(T_{e})$ and $C_{l}$ are the respective heat capacities of
electrons and lattice, $g$ [= $g(T_{l})$] is the $e-ph$ coupling function and
$S(z,t)$ describes the absorbed energy, where $z$ is the depth coordinate. In
the pump-probe experiment $S(z,t)$ has a Gaussian temporal shape. In the limit of
weak perturbation (when change in the electronic temperature is small compared
to the initial temperature) the relaxation of the electronic temperature is
exponential with $e-ph$ thermalization time\cite{Sun93}
\begin{equation}
\tau_{e-ph}=\frac{1}{g}\frac{C_{e}C_{l}}{C_{e}+C_{l}}\text{ \ .} \label{Eq4}
\end{equation}
$g(T_{l})$ is in the linear response limit particularly simple in the case of
simple metals where the electron bandwidth is much larger than the Debye
temperature $\Theta_{D}$, and the Debye model of the electron-phonon coupling
can be used. In this case $g(T)=dG(T)/dT$, where \cite{Groeneveld95,Ahn04}
\begin{equation}
G(T)=4g_{\infty}(\frac{T}{\Theta_{D}})^{5}
{\displaystyle\int\nolimits_{0}^{\Theta_{D}/T}}
\frac{x^{4}dx}{e^{x}-1}\text{ \ .}
\end{equation}
Given that, the T-dependence of $\tau_{e-ph}$ Eq.[\ref{Eq4}] is completely
determined by the $e-ph$ coupling constant ($g_{\infty}$), $\Theta_{D}$, $C_{e}(T)$, and
$C_{l}(T)$. Since the lattice specific heat $C_{l}$ is a factor 10$^{2}$ larger
than the electronic specific heat ($C_{l}$ $\gg$ $C_{e}$) in a wide
temperature range,\cite{Groeneveld95} and since the electron heat capacity
$C_{e}$ = $\gamma$$T_{e} \sim \gamma$$T_{l}$ (when change in the electronic 
temperature is small compared to the initial temperature), the $e-ph$ thermalization 
time $\tau_{e-ph}$ is given by
\begin{equation}
\tau_{e-ph}\approx\frac{C_{e}}{g(T_{l})}=\frac{\gamma T_{l}}{g(T_{l})},
\label{tau}
\end{equation}
For $T_{l}$ $\ll$ $\Theta_{D}$ the function $g(T_{l}$) varies T$_{l}^{4}$ and
for $T_{l}$ $\geq$ $\Theta_{D}$ the function $g(T_{l})$ becomes constant
($g_{\infty}$)\cite{Groeneveld95}. Since $\Theta_{D}=234$ K for Zn, TTM
predicts $\tau_{e-ph}$ $\sim$ T$_{l}^{-3}$ at $T_{l}$ $\leq$ $\Theta_{D}$/5
and $\tau_{e-ph}$ $\sim$ $T_{l}$ at $T_{l}$ $\geq$ $\Theta_{D}$/5.\cite{Ahn04}

\begin{figure}[ptb]
\includegraphics[width=7.5cm]{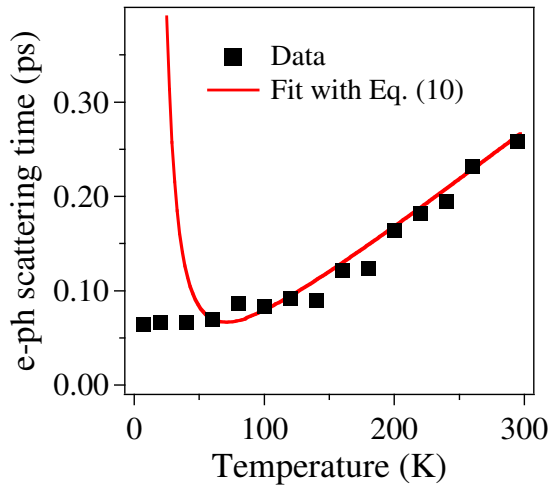}
\caption{(color online) Temperature dependence of
the electron-phonon thermalization time ($\tau_{q}$). The closed squares 
are experimental data, whereas the solid curve is a fit to the data with the TTM model, 
described by Eq. (10).}
\end{figure}
The relaxation time of the subpicosecond transient in Zn obtained by 
fitting the data with Eq. (\ref{Fit}) is shown in Fig. 8 as the function of
lattice temperature. The relaxation time monotonically increases from 75$\pm
$30 fs to 260$\pm$30 fs as the temperature increases, similar to the behavior
obtained for noble metals \cite{Groeneveld95}. The behavior follows the
prediction of the TTM at temperatures above 50K, whereas no upturn in relaxation
time is observed at low temperatures. The absence of upturn in relaxation time
at low temperatures is common in metals,\cite{Groeneveld95} and is attributed
to the fact that the basic assumption of the TTM that $e-e$ thermalization time
is fast compared to the $e-ph$ thermalization fails at low temperatures.
Therefore a nonthermal electron model (NEM) needs to be introduced that
accounts for the low-temperature saturation of relaxation time
\cite{Groeneveld95,Ahn04} - the analysis using a NEM is beyond the scope of
this paper. By fitting the temperature dependence of the relaxation time with
Eq. (\ref{tau}) we obtain the value of the $e-ph$ coupling constant $g_{\infty
}$ = 6.4 x 10$^{16}$Wm$^{-3}$K$^{-1}$, which is the same order of magnitude
as in other simple metals.\cite{Hohlfeld00}

\subsubsection{Temperature dependence of electronic transient}

\begin{figure}[ptb]
\includegraphics[width=8.8cm]{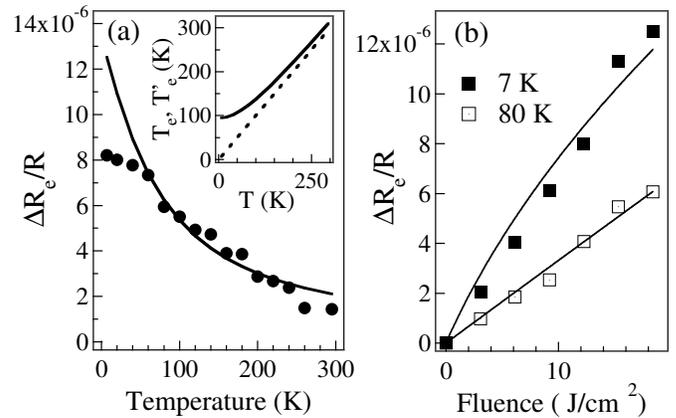}
\caption{(a) Amplitude of the fast electronic transient as a function of temperature. 
Inset shows the corresponding initial ($T_{e}$ - dashed line) and final 
($T_{e}^{\prime}$ - solid line) 
electronic temperatures as a function of temperature. (b) Amplitude of the
fast electronic transient vs the pump fluences at the two typical
temperatures. The solid curves are the fit to the data using Eq. (12). }
\end{figure}
As shown above, the temperature dependence of the relaxation time follows the
prediction of the thermomodulation scenario, where the data above 50 K can be
well fitted using the standard TTM. On the other hand, the amplitude of the
fast electronic response, $B$ in Eq. (\ref{Fit}), also shows pronounced
temperature dependence as shown in Figure 9(a). Moreover the same temperature
dependence is obtained also for the amplitude of the oscillatory transient due
to the photoexcited coherent $E_{2g}$ phonon, suggesting that 
the amplitudes of the fast electronic and the coherent phonon responses 
have the same origin.  

It should be noted that the temperature dependence of the amplitude of the
photoinduced reflectivity transient has thus far not been studied for the case
of simple metals. Below we derive the temperature dependence of the transient
amplitude $B$ as would be expected in the case of the thermomodulation
mechanism and show extremely good agreement with the data.

To account for the temperature dependence of $B$ in Eq. (5), we assume that 
the relaxation processes after excitation with short laser pulse follow the TTM, i.e., 
electron-electron thermalization is much faster than $e-ph$ thermalization.
Furthermore, we assume that the changes in the reflectivity at optical
frequency (1.55 eV) are due to photoinduced absorption, with photoinduced
quasiparticles (in the energy range of $k_{B}T_{e}$ around the Fermi energy)
as initial states and with final unoccupied states well above 
the Fermi energy.\cite{Eesley,Schoenlein87} We should note that similar
arguments have been successfully applied for the case of optical pump-probe
spectroscopy in superconductors \cite{Kabanov} and charge density
waves.\cite{CDW} Therefore, using the Fermi golden rule, the amplitude of the
transient is proportional to the change in the occupied electronic density of
states near $E_{F}$, i.e., the amplitude of the photoinduced transient should
be proportional to the photoinduced quasiparticle density $n_{p}$. In the
limit when temperature is much lower than the Fermi energy, which is the case
here, the number density of the electron-hole pairs $n$ is in the Landau Fermi
liquid exactly proportional to temperature, $n\propto T$. We assume that after
photoexcitation all the absorbed energy initially goes in the electronic
subsystem, and the electrons are described by the increased electronic
temperature $T_{e}^{\prime}$. Following this one obtains 
\begin{equation}
B\propto n_{p}=n_{T_{e}^{\prime}}-n_{T_{e}}\propto T_{e}^{\prime} 
-T_{e}\text{\ \ ,} 
\end{equation}
where $n_{T_{e}}$ is the quasiparticle density at the initial temperature
$T_{e}$, while $n_{T_{e}^{\prime}}$ is the quasiparticle density after
electrons have thermalized to final temperature $T_{e}^{\prime}$. Taking into
account that the electronic specific heat $C_{e}=\gamma T_{e}$, where $\gamma$
is the Sommerfeld constant and using the energy conservation law it follows
\begin{equation}
B\propto T_{e}^{\prime}-T_{e}\text{\ }=\sqrt{T_{e}^{2}+2U_{l}/\gamma} 
-T_{e}\text{ \ ,} \label{Amplitude} 
\end{equation}
where $U_{l}$ is the absorbed energy intensity.

There are several important implications of Eq. (\ref{Amplitude}). First, the
amplitude of the transient is maximum at low temperatures and decreasing\ as
the sample temperature is increasing. Second, at low temperatures the model
predicts a sublinear excitation intensity dependence, while at high
temperatures the dependence becomes more and more linear.

In Fig. 9 we show that the model, indeed, well reproduces the observed
temperature dependence. Here $U_{l}/\gamma$ was taken as a fitting parameter
since the estimated absolute value of $U_{l}$ is subject to some uncertainty
(possible errors in the spot size, reflectance, and the optical penetration
depth may give rise to substantial error). Importantly, the final temperature
$T_{e}^{\prime}$ obtained from the same fit [shown in the inset to Figure 9(a)]
matches well the estimate given in Ref. 23.

Although very good agreement with the temperature dependence of $B$ is obtained
at temperatures higher than $\sim$ 50 K, $B$ shows saturation below 50
K, while the model predicts further increase in the amplitude. We ascribe this
discrepancy to the fact that at low temperatures the TTM fails, since $e-e$
thermalization time becomes comparable or even longer than the $e-ph$
thermalization time.\cite{Ahn04}

Figure 9(b) presents the excitation intensity dependence of amplitude measured
at 7 and 80 K, together with the fit using Eq. (\ref{Amplitude}) using the same
value of $U_{l}/\gamma$ as extracted from the fit to the temperature
dependence data. As seen, the model predicts the sublinear dependence at 7 K,
which is not observed in this intensity range, while good agreement with 80 K
data is obtained. We believe that for the same reasons that the temperature
dependences of relaxation time and amplitude at low temperatures do not follow
the TTM prediction; the fact that electrons are nonthermal on this timescale
can give rise to the absence of the sublinear intensity dependence at low temperatures.

\section{DISCUSSION}

According to the model by Stevens \textit{et al.} \cite{Stevens02} under the condition
that the lattice temperature is constant, the amplitude of the coherent phonon
depends on the imaginary part of the dielectric function Im($\varepsilon$) as
well as the pump intensity $I(t)$ [see Eq. (3)]. Here we review the
temperature dependence of the coherent phonon amplitude in Zn in terms of Eq.
(3). When the lattice temperature increases, thermal expansion occurs and the
volume $v_{c}$ increases. This effect will increase the phonon amplitude
according to Eq. (3). The phonon frequency decreases for $\approx8$ \% with
increasing the temperature from 7 to 295 K, suggesting an amplitude increase
[Eq. (3)] by $\approx$16 \% contrary to our experimental results, which shows
a decrease in the coherent phonon amplitude as the temperature increases.
The deformation potential $\Xi_{0}$ can be taken as nearly constant 
through the entire temperature range in the present study.\cite{Zeiger96} 
Thus, any parameters other than Im($\varepsilon$) in Eq. (3) cannot account for
the temperature dependence of the coherent phonon amplitude.

The initial phase of the coherent optical phonon is estimated to be 5$^{\circ}$ 
$\pm$ 5$^{\circ}$ at all temperatures from the fitting the oscillatory part with 
the single damped harmonic oscillator, indicating that the coherent phonon 
oscillation follows cosine behavior. The 
cosine phase is consistent with the prediction from Stevens' model [see Eq.
(2)], where the driving force of resonant ISRS has displacive character.
Further argument for the displacive nature of the observed transient is the
strong correlation between the photoinduced quasiparticle density and the
amplitude of the coherent optical phonon observed in both temperature and
fluence dependences. For example, in Fig. 5 we use the same fitting function
as used for describing the temperature dependence of the electric transient.
Fitting of the coherent phonon amplitude versus both the temperature and the
fluence using Eq. (12) is surprisingly good. As discussed in Sec. III, the
driving force $F(t)\propto\int_{-\infty}^{t}|E(t^{\prime})|^{2}dt^{\prime}$ is
the Heaviside function convoluted with Gaussian and causes displacive behavior
for the excitation of Raman active mode through resonant ISRS process. Because
Im($\varepsilon$) is directly related to photo excitation of nonequilibrium
electrons, quasiparticle density should follow Im($\varepsilon$): n$_{p}$
$\propto$ Im($\varepsilon$). Thus under the resonant condition, Q$_{0}$ 
$\propto$ Im($\varepsilon$) and n$_{p}$ $\propto$ Im($\varepsilon$). 

\section{CONCLUSIONS}

We have investigated ultrafast dynamics of coherent optical phonons and
nonequilibrium electrons in transition metal Zn and Cd using a femtosecond
pump-probe technique. The optical response in time domain is well fitted by
linear combination of a damped harmonic oscillation due to excitation of
coherent $E_{2g}$ phonon and a subpicosecond electronic transient. Dynamics
of both the nonequilibrium electron distribution and the photoexcited coherent
$E_{2g}$ phonon were observed in a wide temperature range from 7 to 295 K. The
amplitudes of both the coherent optical phonon and electronic transient show
pronounced temperature dependencies. Based on the fact that the relaxation
time of the fast electronic transient closely follows the prediction of the
thermomodulation scenario we analyze the relaxation dynamics in terms of the
two-temperature model. Good agreement with the model is obtained between 50 K
and room temperature, while below 50 K the model is found to fail similar to
studies on simple metals, such as Au or Ag.\cite{Groeneveld95} Below 50 K a
nonthermal electron model \cite{Groeneveld95} needs to be applied. 

In order
to account for the temperature dependence of the amplitude we developed a
model assuming the photoinduced absorption mechanism, where the signal
amplitude is proportional to the photoinduced change in the quasiparticle
density. The model was found to account for the observed temperature
dependence over wide temperature range. Importantly, the model predicts a
sublinear photoexcitation dependence at low temperatures which was not
observed in the excitation range studied. This discrepancy may be due to the
failure of the two-temperature model at low temperatures. 
The fact that the amplitude of the coherent $E_{2g}$ phonon follows the
temperature dependence of the amplitude of the fast electronic transient
suggests that under the resonant condition both electronic and phononic 
responses are proportional to the change in the dielectric function. 

The damping and the frequency of the coherent $E_{2g}$ phonon were also 
found to depend on the lattice temperature, which
was explained by anharmonic phonon-phonon coupling, rather than pure
dephasing, and by anharmonic frequency shift, respectively. There was no
dependence of the frequency and the damping of the $E_{2g}$ phonon on
the pump fluence, suggesting that pure dephasing due to scattering with hot
electrons is negligibly small. When increasing the pump fluence up to
the millijoule per centimeter-squared range, the red-shift of the coherent $E_{2g}$ 
mode frequency was observed, which was ascribed to thermal expansion. 
Since the $E_{2g}$ mode in Zn may be sensitive to highly anisotropic 
compressional behavior of $c/a$ axial ratio in Zn crystal (hcp structure), as observed 
by Raman scattering under pressure, controlling the amplitude and the frequency of 
the coherent $E_{2g}$ phonon will be useful for the study of the gradual anisotropic 
$\rightarrow$ isotropic transition occuring in Zn.\cite{Olijnyk00}

\begin{acknowledgments}
The authors acknowledge Hrvoje Petek and Viktor Kabanov for helpful comments. 
This work was supported by a
Grant-in-Aid for the Scientific Research from the Ministry of Education,
Culture, Sports, Science, and Technology of Japan under grant No. KAKENHI
-15740188 and No. KAKENHI-15035219. The authors acknowledge partly financial 
support from NIMS Research Funds and the Cross-over Research fund of MEXT, Japan.
\end{acknowledgments}

\end{document}